\documentclass[twocolumn,showpacs,preprintnumbers,amsmath,amssymb]{revtex4}
\usepackage{dcolumn}
\usepackage{bm}

\newcommand{\ket}[1]{\displaystyle{|#1\rangle}}
\newcommand{\bra}[1]{\displaystyle{\langle #1|}}
\newcommand{\Er}{\mathbf{E}(\mathbf{r})}
\newcommand{\Br}{\mathbf{B}(\mathbf{r})}
\newcommand{\skj}{\sum_{\mathbf{k}j}}
\newcommand{\skkjj}{\sum_{\mathbf{k}\mathbf{k}'jj'}}
\newcommand{\Ekj}{\mathbf{E}_{\mathbf{k}j}}
\newcommand{\Ekjp}{\mathbf{E}_{\mathbf{k}'j'}}
\newcommand{\akj}{a_{\mathbf{k}j}}
\newcommand{\ackj}{a_{\mathbf{k}j}^\dag}
\newcommand{\akjp}{a_{\mathbf{k}'j'}}
\newcommand{\ackjp}{a_{\mathbf{k}'j'}^\dag}
\newcommand{\fkjr}{\mathbf{f}(\mathbf{k}j,\mathbf{r})}
\newcommand{\ekj}{\mathbf{e}_{\mathbf{k}j}}
\newcommand{\ekjp}{\mathbf{e}_{\mathbf{k}'j'}}
\newcommand{\ukj}{1_{\mathbf{k}j}}
\newcommand{\ukjp}{1_{\mathbf{k}'j'}}

\begin{document}
\preprint{APS/123-QED}
\title{Casimir\,-\,Polder force density between an atom and a conducting wall}
\author{R. Messina and R. Passante}
\email{roberto.passante@fisica.unipa.it}\affiliation{Dipartimento
di Scienze Fisiche ed Astronomiche dell'Universit\`{a} degli Studi
di Palermo and CNSIM,\\Via Archirafi 36, 90123 Palermo, Italy}
\date{\today}
\begin{abstract}
In this paper we calculate the Casimir\,-\,Polder force density
(force per unit area acting on the elements of the surface) on a
metallic plate placed in front of a neutral atom. To obtain the
force density we use the quantum operator associated to the
electromagnetic stress tensor. We explicitly show that the integral
of this force density over the plate reproduces the total force
acting on the plate. This result shows that, although the force is
obtained as a sum of surface element\,-\,atom contributions, the
stress\,-\,tensor method includes also nonadditive components of
Casimir\,-\,Polder forces in the evaluation of the force acting on a
macroscopic object.
\end{abstract}

\pacs{12.20.Ds, 42.50.Ct}
\keywords{Suggested keywords}

\maketitle

\section{Introduction}
A surprising prediction of quantum electrodynamics is the existence
of forces of electromagnetic nature between pairs of metallic
uncharged objects. This kind of effects were first predicted
theoretically by H.B.G. Casimir and D. Polder in $1948$. In two
different articles (the former of Casimir alone \cite{Casimir}, the
latter of both \cite{CasPol}), the existence of forces between two
flat parallel neutral metallic plates, the so called \emph{Casimir
effect}, between a neutral atom and a wall and between two neutral
atoms is predicted. The origin of these forces is commonly
attributed to the properties of the vacuum state of the
electromagnetic field, that is the state of minimum energy, and in
particular to the fact that the vacuum energy depends on the
boundary conditions characterizing the system \cite{Milonni,CPP}.
Although these forces are very tiny, they have been measured with
remarkable precision. The first successful precision experiments
were on the wall\,-\,sphere system \cite{Lamo,MohRoy}. Next, the
force between a neutral atom and a wall was observed
\cite{Sukenik1,Sukenik2,DruzhDeKie}, as well as between two metallic
neutral parallel plates \cite{Bressi1,Bressi2}. For the
atom\,-\,atom Casimir\,-\,Polder force, only indirect evidences of
their existence have been yet obtained.

In this paper we consider a neutral atom in front of a metallic,
perfectly conducting plate. The force on the atom in this
configuration has been calculated by Casimir and Polder
\cite{CasPol}. We focus our attention on the plate and in Section
\ref{SezCalcoli} we calculate, using the quantum electromagnetic
stress tensor, the force acting upon each surface element of the
plate. Then, in Section \ref{SezIntegrale} we show that the integral
of this force density over the surface of the plate equals the
opposite of the force experienced by the atom. This conducts us to a
concluding enquiry on the connection between the stress\,-\,tensor
method and the well\,-\,known nonadditivity of Casimir\,-\,Polder
forces. We conclude that the stress\,-\,tensor method has the great
advantage of automatically including many\,-\,body contributions of
the Casimir\,-\,Polder forces.

\section{The force density on the wall}
\label{SezCalcoli} The Casimir\,-\,Polder interaction between a
neutral atom and a metallic uncharged wall yields an attractive
atom\,-\,wall force. The value of this force was originally
calculated using second order perturbation theory and the minimal
coupling Hamiltonian, obtaining
\begin{equation}
\label{ValoreForza}
F_A(d)=-\frac{3\hbar c\alpha}{2\pi d^5}
\end{equation}
where $d$ is the atom\,-\,wall distance, $\alpha$ is the static
polarizability of the atom and the minus sign indicates that the
force is attractive \cite{CasPol}. The expression
\eqref{ValoreForza} is valid in the so\,-\,called \emph{far zone}
defined by $d>>c/\omega_0$, $\omega_0$ being a typical atomic
frequency.

We now focus our attention on the wall. Being the wall an extended
object, it makes sense to ask ourselves which is the force acting on
each surface element of the metallic plate. To answer this question
we use the quantum operator associated to the classical
electromagnetic stress tensor. This method was widely used by Barton
in his works about fluctuations of Casimir\,-\,Polder forces
\cite{Barton1,Barton2}. In classical electrodynamics, the stress
tensor is defined as
\begin{eqnarray}
\label{DefTensSf}
S_{ij}(\mathbf{r})=\frac{1}{4\pi}
\Bigl[&E_i(\mathbf{r})E_j(\mathbf{r})+B_i(\mathbf{r})B_j(\mathbf{r})\nonumber\\
&-\frac{1}{2}\delta_{ij}\Bigl(E^2(\mathbf{r})+B^2(\mathbf{r})\Bigr)\Bigr]
\end{eqnarray}
where $E_i$ and $B_i$ are components of the electric and magnetic
field, $i$ and $j$ assuming the values $1,2,3$ corresponding,
respectively, to $x$, $y$ and $z$. As it is well known
\cite{Jackson}, this tensor permits to calculate the force on a
volume $V$. The $i$-th component of this force is given by
\begin{equation}
F_i=\int_VdV\,f_i=\oint_S\sum_{j=1}^3dA_j\,S_{ij}
\end{equation}
where the last integral is extended over the surface $S$ enclosing
the volume $V$. If we assume the wall to be located at $z=L$ and
to have an infinitesimal perpendicular extension $dz$, the $z$
component of the force on an infinitesimal parallelepiped $dxdydz$
representing the surface element centered in the point $(x,y)$ on
the plate is given by
\begin{equation}
\label{DensClassica}
\sigma(x,y)=S_{zz}(x,y,L+dz)-S_{zz}(x,y,L)
\end{equation}

Equations \eqref{DefTensSf} and \eqref{DensClassica} are
classical. To obtain the quantum operator associated to the stress
tensor we simply replace the components $E_i$ and $B_i$ of the
electric and magnetic fields with the corresponding quantum
operators. In the Schr\"{o}dinger representation, we have
\begin{equation}
\label{CampoEl}
\Er=i\skj\sqrt{\frac{2\pi\hbar\omega_k}{V}}(\akj-\ackj)\fkjr
\end{equation}
\begin{equation}
\label{CampoMagn}\Br=\skj\sqrt{\frac{2\pi\hbar
c^2}{V\omega_k}}(\akj+\ackj)\bigl[\nabla\times\fkjr\bigr]
\end{equation}
where the functions $\fkjr$ (assumed real) are the field modes
corresponding to the boundary conditions characterizing the
system. In our case, the presence of a metallic surface located in
$z=L$ can be taken into account by considering two metallic boxes
(the former on the left, the latter on the right of the wall)
having in common one side on this plane. At the end of the
calculations, we send to infinity the length of the three edges of
the two cavities. For example, the cavity on the right of the
plane is the parallelepiped
\begin{equation}
-\frac{L_1}{2}<x<\frac{L_1}{2}\quad-\frac{L_1}{2}<y<\frac{L_1}{2}\quad L<z<L_1
\end{equation}
where $L_1>L$ and the volume of the cavity is $V=L_1^2(L-L_1)$.
The mode functions for this box have components
\begin{widetext}
\begin{equation}
\begin{split}f_x(\mathbf{k}j,\mathbf{r})&=\sqrt{8}(\ekj)_x\cos\Bigl[k_x
\Bigl(x+\frac{L_1}{2}\Bigr)\Bigr]\sin\Bigl[k_y\Bigl(y+\frac{L_1}{2}\Bigr)
\Bigr]\sin\Bigl[k_z\Bigl(z-L\Bigr)\Bigr]\\
f_y(\mathbf{k}j,\mathbf{r})&=\sqrt{8}(\ekj)_y\sin\Bigl[k_x
\Bigl(x+\frac{L_1}{2}\Bigr)\Bigr]\cos\Bigl[k_y\Bigl(y+\frac{L_1}{2}\Bigr)\Bigr]
\sin\Bigl[k_z\Bigl(z-L\Bigr)\Bigr]\\
f_z(\mathbf{k}j,\mathbf{r})&=\sqrt{8}(\ekj)_z\sin
\Bigl[k_x\Bigl(x+\frac{L_1}{2}\Bigr)\Bigr]\sin\Bigl[k_y\Bigl(y+\frac{L_1}{2}\Bigr)\Bigr]
\cos\Bigl[k_z\Bigl(z-L\Bigr)\Bigr]\\\end{split}\end{equation}
\end{widetext}
where $\ekj$ are polarization unit vectors and the allowed values
of $\mathbf{k}$ have components
\begin{equation}
k_x=\frac{l\pi}{L_1},\quad k_y=\frac{m\pi}{L_1}, \quad
k_z=\frac{n\pi}{L_1-L},\quad l,m,n=0,1,\dots
\end{equation}
As mentioned before, at the end we take the limit $L,
L_1\to+\infty$.

Since we have replaced the classical stress tensor with a quantum
operator we have to replace the difference on the RHS of
\eqref{DensClassica} with a difference between quantum averages of
the stress tensor operator. These averages must be calculated on
quantum states reflecting the different physical situation at the
two sides of the wall: the bare vacuum on its left side and the atom
on the right side, which we assume located at
$\mathbf{r}_A=(0,0,D)$, with $D>L$. On this basis, we take the
\emph{bare} vacuum state of the electromagnetic field for the space
on the left side of the wall, which we indicate with $\ket{0}$. As
for the right side of the wall, we use the \emph{dressed} vacuum
state, that is the vacuum state corrected by the presence of the
atom. We obtain this state at the lowest significant order in the
atom\,-\,field interaction. It is very convenient to describe the
atom\,-\,field interaction using an effective interaction
Hamiltonian, valid both in the near and the far zone, given by
\cite{PassPowThiru}
\begin{equation}
\label{Interaz}
W=-\frac{1}{2}\skkjj\alpha(k)\Ekj(\mathbf{r}_A)\cdot\Ekjp(\mathbf{r}_A)
\end{equation}
where
\begin{equation}
\Ekj(\mathbf{r})=i\sqrt{\frac{2\pi\hbar\omega_k}{V}}(\akj-\ackj)\fkjr
\end{equation}
are the Fourier components of the electric field \eqref{CampoEl}
and $\alpha (k)$ is the dynamical polarizability of the atom.
Using first order perturbation theory with the interaction
\eqref{Interaz}, we get the dressed vacuum state as
\begin{equation}\label{StatoCorretto}\ket{\tilde{0}}=\ket{0}+\ket{1}\end{equation}
where
\begin{equation}\label{Correzione}\ket{1}=-\frac{\pi}{V}\skkjj\alpha(k)\frac{\sqrt{kk'}}{k+k'}\mathbf{f}(\mathbf{k},j,\mathbf{r}_A)\cdot
\mathbf{f}(\mathbf{k}',j',\mathbf{r}_A)\ket{\ukj\ukjp}\end{equation}
$\ukj$ denoting the presence of a photon with wavevector
$\mathbf{k}$ and polarization $j$.

Thus, the force density is expressed by
\begin{equation}
\label{DensQuant}
\sigma(x,y)=\bra{\tilde{0}}S(x,y)\ket{\tilde{0}}-\bra{0}S(x,y)\ket{0}
\end{equation}
where, for simplicity of notations,
\begin{equation}
S(x,y)=S_{zz}(x,y,L).
\end{equation}

The explicit expression of the operator $S(x,y)$ can be simply
obtained from \eqref{DefTensSf} in the following form
\begin{widetext}
\begin{equation}
\label{TensSforzi}
\begin{split}
S(x,y)=-\frac{2\hbar c}{V}\skkjj\Biggl\{&A(\mathbf{k}j)A(\mathbf{k}'j')(\akj-\ackj)(\akjp-\ackjp)+\\
&+B(\mathbf{k}j,\mathbf{k}'j')(\akj+\ackj)(\akjp+\ackjp)\Biggr\}.\\
\end{split}
\end{equation}
where
\begin{equation}
A(\mathbf{k}j)=\sqrt{k}(\ekj)_z\sin\Bigl[k_x\Bigl(x+\frac{L_1}{2}\Bigr)\Bigr]
\sin\Bigl[k_y\Bigl(y+\frac{L_1}{2}\Bigr)\Bigr]
\end{equation}
and
\begin{equation}
\begin{split}
B(\mathbf{k}j,\mathbf{k}'j')=\frac{1}{\sqrt{kk'}}\Biggl\{
&\Bigl((\ekj)_zk_x-(\ekj)_xk_z\Bigr)\Bigl((\ekjp)_zk_x'-(\ekjp)_xk_z'\Bigr)\cdot\\
&\cdot\cos\Bigl[k_x\Bigl(x+\frac{L_1}{2}\Bigr)\Bigr]\cos\Bigl[k_x'
\Bigl(x+\frac{L_1}{2}\Bigr)\Bigr]\sin\Bigl[k_y\Bigl(y+\frac{L_1}{2}
\Bigr)\Bigr]\sin\Bigl[k_y'\Bigl(y+\frac{L_1}{2}\Bigr)\Bigr]+\\
&+\Bigl((\ekj)_zk_y-(\ekj)_yk_z\Bigr)\Bigl((\ekjp)_zk_y'-(\ekjp)_yk_z'\Bigr)\cdot\\
&\cdot\sin\Bigl[k_x\Bigl(x+\frac{L_1}{2}\Bigr)\Bigr]\sin\Bigl[k_x'
\Bigl(x+\frac{L_1}{2}\Bigr)\Bigr]\cos\Bigl[k_y\Bigl(y+\frac{L_1}{2}\Bigr)\Bigr]
\cos\Bigl[k_y'\Bigl(y+\frac{L_1}{2}\Bigr)\Bigr]\Biggr\}.\\
\end{split}
\end{equation}
\end{widetext}

Using eq. \eqref{StatoCorretto}, from \eqref{DensQuant} we obtain
at the first order in $\alpha$,
\begin{equation}
\label{S01}
\sigma(x,y)=2\bra{0}S(x,y)\ket{1}=2S_{01}(x,y).
\end{equation}

Using \eqref{Correzione} and \eqref{TensSforzi}, we obtain
\begin{equation}
\begin{split}
\sigma(x,y)&=\frac{8\pi\hbar
c}{V^2}\skkjj\alpha(k)\frac{\sqrt{kk'}}{k+k'}\mathbf{f}(\mathbf{k}j,\mathbf{r}_A)\cdot
\mathbf{f}(\mathbf{k}'j',\mathbf{r}_A)\times\\
&\times\Bigl[A(\mathbf{k}j)A(\mathbf{k}'j')+B(\mathbf{k}j,\mathbf{k}'j')\Bigr].
\end{split}
\end{equation}

Our system has a cylindrical symmetry around the axis
perpendicular to the wall and passing through the atom. Thus,
being $\mathbf{P}=(0,0,L)$ the point common to this axis and the
wall, the force density depends only on the distance $\rho$ of a
point of the plate from $\mathbf{P}$. In the far zone, where the
dynamical polarizability $\alpha(k)$ is replaced with its static
value, we obtain
\begin{equation}
\sigma(\rho)=\frac{\hbar c\alpha}{4\pi^3}\int_0^{+\infty}dx\,
\Bigl[I_1^2(x)+2I_2^2(x)+I_3^2(x)+I_4^2(x)\Bigr]
\end{equation}
where
\begin{widetext}
\begin{equation}
\begin{split}
I_1(x)&=\int_0^{+\infty}dk\,k^3e^{-kx}\int_0^\pi d\theta\,\sin^3\theta\cos
\bigl(kd\cos\theta\bigr)J_0\bigl(k\rho\sin\theta\bigr)\\
I_2(x)&=\int_0^{+\infty}dk\,k^3e^{-kx}\int_0^\pi d\theta\,
\sin\theta\cos\theta\sin\bigl(kd\cos\theta\bigr)J_0\bigl(k\rho\sin\theta\bigr)\\
I_3(x)&=\int_0^{+\infty}dk\,k^3e^{-kx}\int_0^\pi d\theta\,
\sin^2\theta\cos\bigl(kd\cos\theta)J_1\bigl(k\rho\sin\theta\bigr)\\
I_4(x)&=\int_0^{+\infty}dk\,k^3e^{-kx}\int_0^\pi
d\theta\,\sin^2\theta\cos\theta\sin\bigl(kd\cos\theta\bigr)J_1(k\rho\sin\theta\bigr)\\\end{split}
\end{equation}
\end{widetext}
where $J_\nu(x)$ is a Bessel function of the first kind of order
$\nu$ \cite{Abram}. Making use of known properties of Bessel
functions \cite{Bessel}, we finally obtain the following
expression for the force density
\begin{equation}
\label{RisDensFor} \sigma(\rho)=\frac{\hbar
c\alpha}{4\pi^2}\,\frac{17d^2+10\rho^2}{(d^2+\rho^2)^{\frac{9}{2}}}\end{equation}
where $d=D-L$ is the atom\,-\,wall distance. It is immediate to
see that the force density is vanishing in both limits
$d\to+\infty$ (atom infinitely distant from the wall) and
$\rho\to+\infty$ (surface element of the wall infinitely distant
from the atom).

\section{The integral of the force density and nonadditivity of Casimir\,-\,Polder forces}
\label{SezIntegrale} Once the force density \eqref{RisDensFor} has
been obtained, we can integrate it over the surface of the wall in
order to obtain the total force $F_W(d)$ experienced by the wall.
We easily obtain
\begin{equation}
F_W(d)=2\pi\int_0^{+\infty}d\rho\,\rho\,\sigma(\rho)=\frac{3\hbar
c\alpha}{2\pi d^5}
\end{equation}
which is the opposite of the force $F_A(d)$ acting on the atom
given by equation \eqref{ValoreForza}.

This result may appear contradicting the well\,-\,established fact
that Casimir\,-\,Polder forces are not additive
\cite{Milonni,PassPers,CirPass}. For example, in the case of three
atoms it is known that the force on one of them is not simply the
sum of the forces due to the other two atoms separately. The
system we are considering, and in particular the fact that we are
focusing our attention on elements of the plate, raises a similar
problem. Suppose we have the atom in front of a single element of
the plate and that we calculate the force acting on this surface
element. This force, obviously, will depend on the coordinate
$\rho$ of the element. Due to the nonadditivity of
Casimir\,-\,forces, we expect that its integral over the entire
plate should \emph{not} give the total force, since the force on
the wall should also contain three\,-\,body components involving
the atom and two different elements of the wall (all these
components are proportional to $\alpha$). The solution of this
seemingly contradictory point is related to the use of the
stress\,-\,tensor method. The stress tensor operator
\eqref{DefTensSf} contains in its very expression the electric and
magnetic field operators. These field operators, given by
equations \eqref{CampoEl} and \eqref{CampoMagn}, involve the mode
function $\fkjr$, which are taking into account the presence of
the entire plate. As a consequence, we claim that the use of the
operator $S(x,y)$ to calculate the force density on a surface
element of the wall answers the following question: if we have a
neutral atom in front of a conducting wall, what is the force
acting on a surface element of the wall in the presence of the
\emph{entire} plate? On the basis of this consideration, the
integral of such contribution over the wall surface must give the
correct value for the total force, as we have explicitly shown.
Hence, all many\,-\,body contributions to the atom\,-\,surface
element Casimir\,-\,Polder force which are proportional to
$\alpha$ are already included in our result \eqref{RisDensFor}.
Moreover, we wish to stress that from the density of the force we
can obtain much more information on the effects of the
atom\,-\,wall Casimir\,-\,Polder interaction, such as torques or
stresses on the wall due to the presence of the atom. These
effects may be relevant in the recently proposed technological
application of Casimir forces \cite{Lamoreaux}.

\section{Conclusions}
In this paper we have considered the Casimir\,-\,Polder
interaction between a neutral atom and a neutral conducting wall.
We have calculated, using the quantum operator associated with the
classical electromagnetic stress tensor, the force density on the
plate, that is the force acting on each surface element of the
wall. We have shown that the integral of this density gives the
correct result for the total forced acting on the plate, i.e. the
opposite of the force on the atom. This shows that our method
based on the stress tensor, usually used for macroscopic bodies,
enables to include easily, at the order considered, all
nonadditive components of Casimir\,-\,Polder forces between the
atom and two or more plate elements. In fact, the force calculated
for each surface element is indeed the force acting on it in the
presence of the entire plate. This happens because the stress
tensor in its very definition contains the modes of the
electromagnetic field, which take into account the presence of all
macroscopic bodies in the system.

\begin{acknowledgments}
This work was in part supported by the bilateral Italian\,-\,Belgian project on
"Casimir\,-\,Polder forces, Casimir effect and their fluctuations"
and by the bilateral Italian\,-\,Japanese project 15C1 on "Quantum
Information and Computation" of the Italian Ministry for Foreign
Affairs. Partial support by Ministero dell'Universit\`{a} e della
Ricerca Scientifica e Tecnologica and by Comitato Regionale di
Ricerche Nucleari e di Struttura della Materia is also
acknowledged.
\end{acknowledgments}

\end{document}